# Neutron reflection from the surface of normal and superfluid $^4$He

T. R. Charlton[1], R. M. Dalgliesh[1], A. Ganshin[2], O. Kirichek[1], S. Langridge[1], P. V. E. McClintock[2]

[1] *ISIS, STFC, Rutherford Appleton Laboratory, Harwell, Didcot, UK*
[2] *Physics Department, Lancaster University, Lancaster, LA1 4YB, UK*

**The reflection of neutrons from a helium surface has been observed for the first time. The $^4$He surface is smoother in the superfluid state at 1.54 K than in the case of the normal liquid at 2.3 K. In the superfluid state we also observe a surface layer ~200 Å thick which has a subtly different neutron scattering cross-section, which may be explained by an enhanced Bose-Einstein condensate fraction close to the helium surface.**

The importance of liquid helium for physics can hardly be overestimated. Its unique ability to remain liquid right down to absolute zero makes helium an ideal experimental sample for studying quantum phenomena in condensed matter. Helium-4 was first liquefied by Kamerlingh Onnes in 1908. Since then the properties of bulk helium liquid (including $^4$He, $^3$He and isotopic mixtures) have been studied in detail, resulting in the discovery of many fundamental phenomena including, especially, superfluidity. However the free surface of liquid helium has attracted significantly less attention than the bulk liquid, perhaps due to the relatively complex experimental methods that are needed for surface studies. Currently the best-known approaches are probably optical ellipsometry, surface tension measurements, and investigations of two-dimensional charge sheets above and below the surface.

Optical ellipsometry was exploited for measuring the thickness and density profile of the liquid helium surface [1]. It was shown in experiments that the vertical thickness within which the density changes from 90 to 10% of the bulk liquid density increases slightly with temperature from 1.4 to 2.1 K , with an average value of 9.4 Å at 1.8 K. A very close value for the vertical thickness was obtained in [2].

The surface tension of liquid helium can be measured by various experimental methods [3, 4] and may be compared directly with the values derived from theoretical models [5, 6]. However, establishing the relationship between surface tension and the dynamical properties of superfluid helium remains a rather complex task [7].

Perhaps the most powerful experimental technique for studying the surface properties of liquid helium is through measurements on two-dimensional charge systems including surface electrons (SE) and layer of ions trapped under the surface of liquid helium [8]. The existence of the quantized capillary surface waves, or ripplons, earlier proposed by Atkins [7] in order to explain surface tension data, has been clearly proven in experiments with a two-dimensional plasma resonance in SE [9] and in complete control conditions [10]. Later, ripplons [11] as well as scattering of SE with Fermi quasi-particles from bulk liquid [12] were detected on liquid $^3$He. Soon after that, it was found that the mobility of SE on $^3$He deviates drastically from the single-electron-ripplon scattering theory below 70 mK [13]. In contrast, the mobility of SE on $^4$He follows the single-electron-ripplon scattering theory down to 20 mK, the lowest temperature achieved in the experiment. This may suggest a crossover to the long-mean-free-path regime, implying a connection to the Rayleigh-like waves predicted theoretically in [14, 15]. A peculiar temperature dependence was also observed during studies of the surface tension of liquid $^3$He [4], behavior that could not be explained in terms of ripplons and the quasi-particle-scattering contribution.

Capillary waves on the surface of thick superfluid $^4$He films have been detected via the dielectric ponderomotive effect [16]. This method allowed measurement of the spectrum and damping of capillary waves within the wavelength range 3.3 – 20 μm. The damping was orders of magnitude smaller than the value predicted theoretically.

In some respects, the SE, surface tension and capillary wave methods provide rather indirect ways of studying the liquid helium surface experimentally, inevitably leading to ambiguities in the interpretation of the data. A complementary experimental approach that might facilitate direct studies of the microscopic properties of liquid helium surface, thus illuminating this fascinating subject, is therefore much to be desired.

In this paper we propose a new experimental method based on the unique combination of neutron reflection and ultra-low temperature sample environment: small-angle neutron reflection from the liquid surface, opening new opportunities for studying the interface properties of quantum liquids. We also present our preliminary measurements of reflection from the superfluid and normal liquid $^4$He surfaces.

Total reflection of slow neutrons was first reported by Fermi and co-workers [17, 18] and since been extensively applied to range of problems. The specular reflection of neutrons gives information on the neutron refractive index profile normal to a surface. The refractive index is simply related to the scattering length density and hence specular neutron reflection can provide important information about the composition of surfaces and interfaces.

Our experiments were performed on the CRISP Instrument, which was designed as a general purpose reflectometer for the investigation of a wide spectrum of interfaces and surfaces. It uses a broad-band neutron time-of-flight method for determination of the wavelength at fixed angles. The sample geometry is horizontal, allowing investigation of liquid surfaces. The instrument uses the pulsed neutron beam from the ISIS neutron facility and passage via a 20 K hydrogen moderator provides neutron wavelengths in the range 0.5 - 6.5 Å at the source frequency of 50 Hz.

For the experiment to be described, we used a Variox $^{BL}$ cryostat for neutron scattering experiments with a variable temperature insert. A cylindrical cell made of aluminium alloy 6082 with inner diameter 31.4 mm, internal length 74.0 mm and wall thickness 0.5 mm was attached to the sample stick of the insert.

In order to estimate the influence of the cell and sample environment on the signal from the scattered neutrons, we performed neutron reflection calibration measurements from the surface of liquid D$_2$O at room temperature in the same experimental setup. The test demonstrated a negligible influence of the sample environment on the experimental data.

In our experiments we attempted to study neutron reflection from the surfaces of both liquid $^4$He and $^3$He. However, due to the large cross-section for neutron scattering from the $^3$He nucleus in the vapour phase, we could not detect a reflection signal from the $^3$He liquid surface within the experimental temperature range down to 1.25K. The $^3$He vapour pressure decreases rapidly with temperature, and therefore we believe that the temperature reduction below ~1K should solve this problem.

In the experiment we condense approximately 10 litres of helium gas (at STP) into the cell, which is kept at a temperature of 2 K. After the reflection signal is detected, the temperature of the cell is stabilized and controlled for a sufficient period of time to collect neutron reflection data over a range of values of wave vector transfer $Q_z$ perpendicular to the reflecting surface (see the diagram in the inset of Fig. 1), which typically takes a few hours.

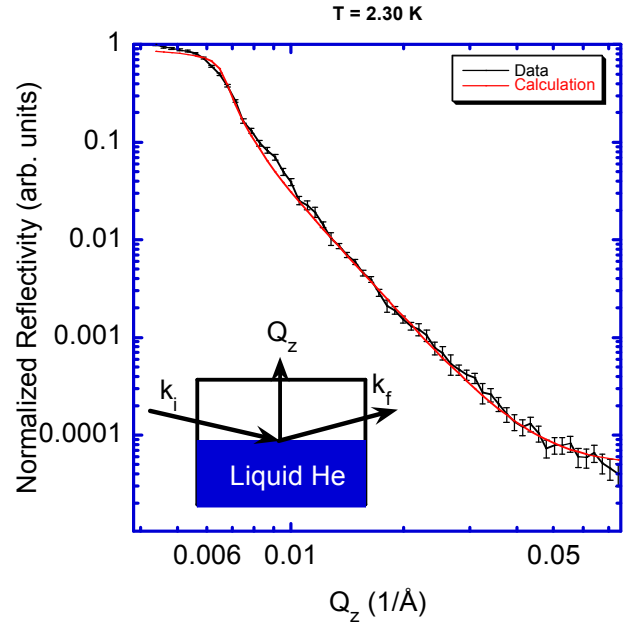

Fig. 1 The reflectivity of the liquid $^4$He surface as a function of $Q_z$ measured at 2.3 K. The inset shows the orientation of the incoming $k_i$ and outgoing $k_f$ wave vectors as well as the momentum transfer vector $Q_z$ in relation to the liquid helium surface.

We have measured the reflectivity of the liquid $^4$He surface as a function of $Q_z$ at temperatures of 2.3 K and 1.54 K, obtaining the results presented in Figs. 1 and 2 respectively. As a general feature both data sets show a slight downward slope of the reflectivity approaching the critical value, $Q_c$~0.006 1/Å, attributable to absorption of neutrons by traces of $^3$He at the natural isotopic ratio of about $2 \times 10^{-7}$ in the supplied $^4$He gas. Without absorption the signal below $Q_c$ would be completely flat. Past $Q_c$, from $Q_z >$ 0.08 1/Å where the Born approximation applies we observe the typical ~ $Q_z^{-4}$ decrease in the reflectivity curve indicating, qualitatively, that the surface is smooth.

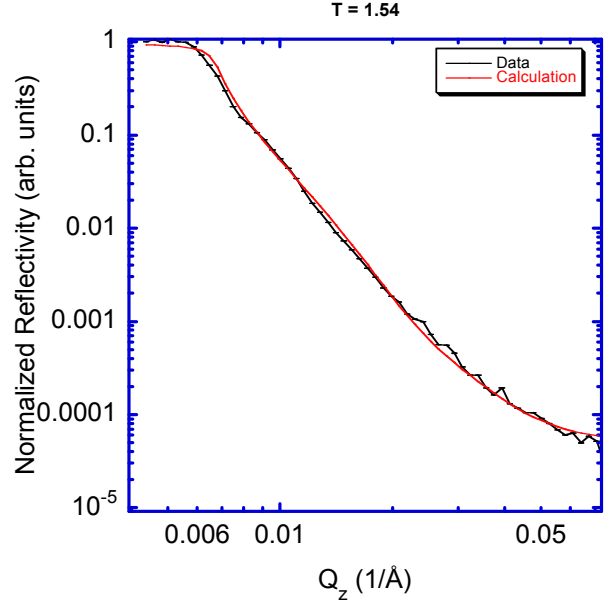

Fig. 2 The reflectivity of liquid $^4$He surface as a function of $Q_z$ measured at 1.54 K.

To extract quantitative information from the data we compare the reflectivity profile with an optical model. In its simplest form, the model reduces to the quantum mechanical problem of a particle incident on a potential step. In our case, the optical potential has three distinct regions (vacuum, near surface, and bulk) with the transition between adjacent steps calculated as an interface roughness. The reflectivity is calculated from the optical potential using a recursive definition of the reflectivity, taking into account multiple reflections [19]. The numerical description of the interface roughness follows that of reference [20]. The results of the calculation are shown as lines through the data in Fig. 1 and 2 with the optical potential shown in Fig. 3. From the calculations it is clear that, near the surface, a region with higher scattering power exits in the 1.54K case as compared to the 2.30K calculations. Also the transition from bulk to vacuum at 2.30K occurs gradually over a large 800 Å distance while, at 1.54K, the transition is compressed into ~400 Å. We can suggest a possible explanation for this observation.

Experiments on liquid $^4$He with both neutron scattering [21, 22] and X-ray diffraction [23] indicate a significant change in the structure of liquid as it cools through the superfluid transition. This observation was also supported by Monte Carlo simulations [24] where it was shown that, when Bose condensation takes place, the liquid $^4$He becomes more disordered in Cartesian space with some rearrangement of atoms occurring to create the necessary space for delocalization to occur. If there is an enhanced Bose-Einstein condensate fraction close to the surface of the superfluid [25, 26], an increase in disorder could be expected, possibly accounting for the alteration in the neutron scattering structure factor. However thorough theoretical modelling is now required to establish whether the results observed in the experiment can be accounted for quantitatively on this basis.

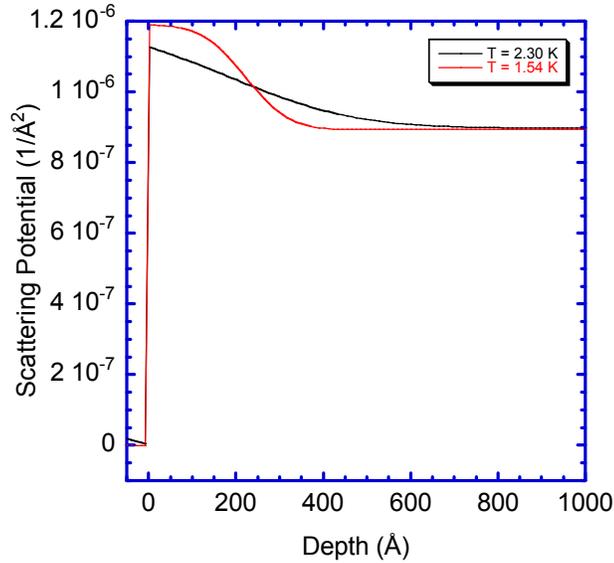

Fig. 4 The scattering potential profiles obtained by numerical simulation, using the reflectivity data for 2.3 K and 1.54 K. The abscissa corresponds to vacuum below 0 Å, to the near-surface region between 0 Å and 800 Å, and to bulk liquid $^4$He above 800 Å. Note that the optical potential is proportional to the density and scattering power of the material.

In conclusion, we have observed neutron reflection from the $^4$He surface for the first time. We also have observed a surface layer ~200 Å thick with a subtly different neutron scattering cross-section in superfluid $^4$He. The new experimental method described in this paper opens up fresh opportunities in the study of quantum fluids and solid quantum surfaces and interfaces. For example it could be used for direct observation of Fomin surface excitations on liquid $^3$He [14], or in the search for a superfluid transition in 2D $^3$He on the surface of micro-separated $^3$He-$^4$He liquid mixture [27].

We gratefully acknowledge help from the ISIS sample environment group.